# Exploring the relationship between built environment and bike-sharing demand: Does the trip length matter?

Feiyang Wang[a†], Chaoying Yin[b†], Ximing Chang[c], Der-Horng Lee[d], Zhengbing He [a*]

a. Beijing Key Laboratory of Traffic Engineering, Beijing University of Technology, Beijing, China
b. College of Automobile and Traffic Engineering, Nanjing Forestry University, Nanjing, China;
c. State Key Laboratory of Rail Traffic Control and Safety, Beijing Jiaotong University, Beijing, China
d. ZJU-UIUC Institute, Zhejiang University, Zhejiang, China

† Equal contribution
* Corresponding author: he.zb@hotmail.com

**Abstract:** Bike-sharing has received considerable practice and research attention over the past decade. As a manpower-driven transportation mode, it seems more sensitive to trip length, since one could take a shared bike to a destinated place where is too far to walk, or choose it for simply replacing walking when going to a nearby place. However, little research has paid attention to it, i.e., the differentiated effects of built environment on the bike-sharing demand with trip lengths. To fill the gap, this paper identifies a threshold of bike-sharing trip lengths from bike-sharing trace data, and employs a semiparametric geographically weighted Poisson regression (SGWPR) model to investigate the relationship between built environment and bike-sharing demand with different lengths considering the heterogeneity in the relationship. The results show that built environment has heterogeneous effects on the bike-sharing demand in urban areas, and the effects differ across groups with trip lengths. The findings contribute to understanding the relationships between built environment and bike-sharing demand, and providing supports for the placements and dispatchment of shared bikes.

# 1 Introduction

In recent years, bike-sharing has received increasing attention with the development of Internet communication technology and mobile payment. Taking China as an example, more than 19.5 million shared bikes have been put into use in 360 Chinese cities, which covered 300 million people (Eren and V. E. Uz, 2020). Bike-sharing is viewed as a newly-emerging green transportation mode, due to the flexible, healthy, low-carbon and environment-friendly features (Barbour et al., 2019; Li et al., 2021). Bike-sharing has effectively solved first- and last-mile problems to some extent (Liu and Lin, 2019), helping mitigate traffic congestion and air pollution.

Although the literature has extensively investigated the impact factors of bike-sharing demand (Eren and V. E. Uz, 2020; Wang et al., 2020; Ji et al., 2018), there are still two main research gaps. *First*, most of the literature focuses on the relationship between built environment and bike-sharing demand by directly analyzing origins and destinations of bike-sharing (Gao et al. 2021; Shen at al., 2018; Chang at al., 2020). In reality, different travel purposes are usually associated with different trip lengths (Radzimski et al., 2021). For example, the travel distances are usually short when complementing public transit travels, and thus short-distance riding trips tend to concentrate around core urban areas where better public transportation services are provided. By contrast, long-distance riding trips show substitution effects on public transit travels (Levy et al., 2019). Therefore, it is necessary to understand the relationship between built environment and bike-sharing demand with trip lengths. However, location-based origin and destination data of bike-sharing, which are commonly used in the literature, cannot provide accurate trip lengths because cyclists usually choose cyclist-friendly routes, but not the shortest ones (Lu et al., 2018). *Second*, most existing studies mainly focus on the bike-sharing demand around public transit stations or bike-sharing stations, establishing the relationship between built environment and bike-sharing demand from a local perspective (El-Assi et al., 2017; Guo and He, 2020; Wu et al., 2021). However, these studies cannot provide the global mobility patterns of bike-sharing that is important for the scheduling and planning of bike-sharing. Moreover, few studies have addressed the spatiotemporal heterogeneity in the relationship between built environment and bike-sharing demand.

To fill the above gaps, this paper employs a semiparametric geographically weighted Poisson regression (SGWPR) model to explore the heterogeneous relationship between built environment and bike-sharing demand based on Global Navigation Satellite System (GNSS)-based trace data from Tianjin, China. This paper contributes to the literature by: (1) identifying a threshold of bike-sharing trip lengths and investigating the relationship between built environment and bike-sharing demand with different trip lengths; (2) addressing the spatiotemporal heterogeneity in the relationship and understanding the global mobility patterns by dividing the urban areas into same-size square grids.

The organization of the rest of the paper is as follows. Section 2 provides a brief review on the existing studies of the determinants of bike-sharing. Section 3 introduces the data used and variables selected in this study. Sections 4 and 5 describe the research method and estimated results, respectively, and Section 6 presents the key findings and discusses potential policy implications.

## 2 Literature review

### 2.1 Understanding the bike-sharing demand

A growing body of research is dedicated to understanding the bike-sharing demand, in which multiple data sources are used. Traditional questionnaire survey collects detailed travel information of bike-sharing users, and it has become one of the most common data sources. For example, Xin et al. (2018) conducted a user satisfaction survey in Shanghai, China, and found that company employees are the major bike-sharing users, followed by college students and self-employed workers. Based on survey data from Washington DC and Minneapolis, the United States, Martin et al. (2014) found that travelers could switch toward or from public transit as an outcome of bike-sharing and the effects varied across their residential locations. Based on stated preference survey data, Mohanty et al. (2017) analyzed the relationship between bike-sharing demand and transit ridership. The results show a positive relationship and no significant heterogeneity among the surveyed sample. Based on survey data from Ningbo, China, Guo et al. (2017) adopted a bivariate ordered probit model to explore the factors of bike-sharing demand and satisfaction with this mode, and found that most socio-economic characteristics (e.g. income) are determinants.

Except for survey data, emerging big data (e.g., origin and destination data of bike-sharing) have been used to explore bike-sharing usage behavior, which can better uncover bike-sharing mobility. For example, using the origin and destination data of bike-sharing from Singapore, Shen et al. (2018) applied a spatial autoregressive model to explore the spatiotemporal mobility patterns of bike-sharing, in which the impacts of built environment, public transit, bike infrastructure and weather were considered. The results of the spatial autoregressive model suggest that more balanced land use, better public transit accessibility and sufficient bike supporting facilities can encourage bike-sharing usage. Based on spatial statistics and graph-based methods, Yang et al. (2019) found that new metro services promoted bike-sharing demand around metro stations and the impacts had spatial variations. Li et al. (2020) used non-negative matrix factorization to capture bike-sharing usage patterns and explored the impacts of built environment and socio-economic characteristics on bike-sharing demand by using the geographically weighted regression (GWR) model. The results confirm the spatial variations in the impacts. Gao et al. (2021) explored the distance decay in bike-sharing usage based on the origin and destination location data and a multiple linear regression model was applied to explore the link between the distance decay and built environment. The results show that there exist spatial heterogeneities in the link. Yu et al. (2021) explored the spatiotemporal mobility patterns of feeder bike-sharing of metro and found that the demand of shared bikes follows a power-law distribution. Moreover, there exist differences in the bike-sharing demand on weekdays and weekends.

GNSS-based trace data of bike-sharing is the data that is commonly used to analyze bike-sharing demand. GNSS-based trace data contains locations of each shared bike that are updated with a fixed interval. Thus, GNSS-based trace data can provide more accurate moving trajectories. Based on GNSS-based trace data, Lu et al. (2018) found that bike-sharing users were more likely to choose environmentally friendly routes, but not the shortest ones. Ji et al. (2020) explored the usage patterns and determinants of docked and dockless bike-sharing and confirmed the difference between two types of bike-sharing systems.

**2.2 Models in bike-sharing related studies**

In the literature, various methods have been applied to investigate the impact factors of using bike-sharing, such as discrete choice models, spatial clustering models and machine learning models. For example, Liu et al. (2019) used a multinomial logit model to explore the association between land use and bike-sharing demand from a spatiotemporal perspective. Guo et al. (2020) applied a negative regression model to explore the impacts of built environment on feeder bike-sharing of metro and found a negative relationship. Based on statistical methods, these studies confirm the impacts of built environment and socio-economic characteristics on bike-sharing demand. Focusing on the bike-sharing demand around, Central Business District and Beijing West Railway Station, Wang et al. (2022) used a random forest method to investigate the determinants of bike-sharing demand and found that built environment shows nonlinear effects on bike-sharing demand. However, the impacts may show spatial variations owing to the spatial heterogeneities. Qian and Ukkusuri (2015) used the GWR model to explore the spatial variation of the taxi demand, in which the GWR model was shown to be a solution to address the spatial heterogeneities. To address the spatial heterogeneities, Wang et al. (2020) used a GWR model to explore the impacts of metro passenger flow and built environment around metro stations on bike-sharing demand. In another example, a geographically weighted Poisson regression (GWPR) model is used to explore the impacts of built environment on metro-bikeshare transfer flow based on smart-card data of metro and bike-sharing (Ji et al., 2018). These studies confirm the necessity of addressing the spatial heterogeneities when modeling bike-sharing demand.

**2.3 Comments on the existing research**

Two categories of data (e.g., survey data and GNSS data of bike-sharing) are the main data sources for the exploring of the impacts of built environment on bike-sharing demand. However, they have the following limitations. Survey data cannot provide real-world behaviors of bike-sharing users although detailed information of individual characteristics is included. In addition, the sample of survey data tends to be small. Emerging big data, such as GNSS-based data of bike-sharing, can promise enough samples for analyzing the bike-sharing demand. However, most existing studies usually GNSS-based origin and destination data to explore the relationship between built environment and bike-sharing demand. The GNSS-based origin and destination data of bike-sharing limit the potential to analyze bike-sharing demand from a length perspective as they cannot provide accurate trip lengths.

Moreover, most existing studies focus on bike-sharing demand that is close to public transit stations or bike-sharing stations. Although most short-distance bike-sharing trips are used as a feeder of public transit, long riding trips show substitution effects on public transit travel. Thus, these studies cannot provide a global portrait of bike-sharing demand with spatiotemporal heterogeneity.

To overcome the limitations, GNSS-based trace data of bike-sharing are analyzed in this study. Using the high-fidelity trace data, trip lengths are calculated and a reasonable threshold of trip lengths is identified to divide the bike-sharing demand. By dividing the urban areas into square grids, this study employs a SGWPR model to explore the spatiotemporal heterogeneity in the relationship between built environment and bike-sharing demand with different trip lengths.

## 3 Data description

Tianjin, which is one of four municipalities, is one of the largest cities in China. Six central urban districts in Tianjin are selected as research areas, as shown in Figure 1(a). To simplify the district names, *Heping* district, *Hebei* district, *Hongqiao* district, *Nankai* district, *Hexi* district, *Hedong* district are represented with A, B, C, D, E and F, respectively. Three expressways, namely *Inner Ring*, *Middle Ring* and *Outer Ring*, enclose these districts. The lengths of three expressways are 16 km, 35 km and 50 km, respectively. In this study, the bike-sharing demand is analyzed based on a unit of same-size square grids (500 m × 500 m) and 728 grids are obtained (Figure 1(b)). The grids help to reduce interferences of error data, compared with the commonly used blocks.

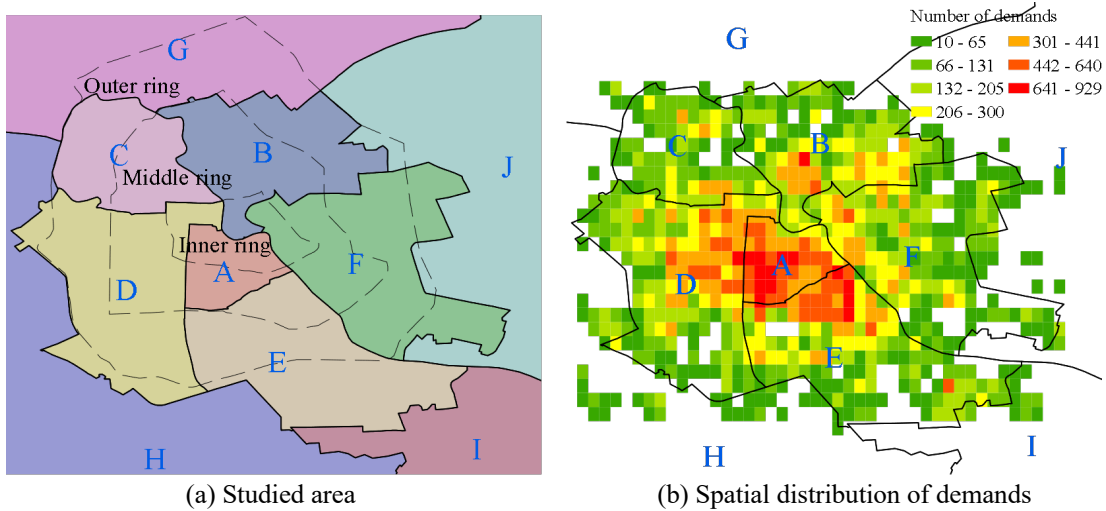

(a) Studied area　　　　(b) Spatial distribution of demands

Figure 1 Research areas and grid-based bike-sharing demand of Tianjin, China.

The GNSS-based trace data are provided by one of the biggest bike-sharing operators. To better understand the spatiotemporal demand, we clean the raw data and remove those records of bike-sharing with the following characteristics: (1) trips longer than 15 km or shorter than 0.1 km; (2) average riding speed faster than 35 km/h; and (3) the origin or destination beyond research areas. Bike-sharing trips generated on May 6-10 and May 13-17, 2019 are used for this analysis. Considering the significant differences in bike-sharing demand between weekdays and weekends, we focus on the bike-sharing demand on only weekdays in this study and 2.22 million valid trips are obtained for this analysis. The cumulative probability distribution of trip lengths is shown in Figure 2. This study aims to investigate the bike-sharing demand with different trip lengths. A sensitivity analysis is conducted to identify a rational threshold of bike-sharing trip lengths. The results of sensitivity analysis confirm that 1.5 km is an appropriate threshold to subdivide the database, which is consistent with Radzimski et al. (2021)'s study.

The process and discussion of the sensitivity analysis will be described in detail in Section 5.2.1. Thus, the trips are divided into two categories in this study: (1) short trips, whose lengths are between 0.1 km and 1.5 km, and (2) long trips, whose lengths are longer than 1.5 km. In the database, short trips account for 60% of the total trips.

Figure 3 presents the hourly changes of bike-sharing demand and it suggests that there are two peaks in bike-sharing demand and the peak hours are approximately 8:00 and 18:00, respectively. Therefore, two peaks (7:00-9:00 and 17:00-19:00) and an off-peak (11:00-13:00) are analyzed, respectively. The bike-sharing demand at different periods, i.e., the dependent variables of this study, is shown in Table 1.

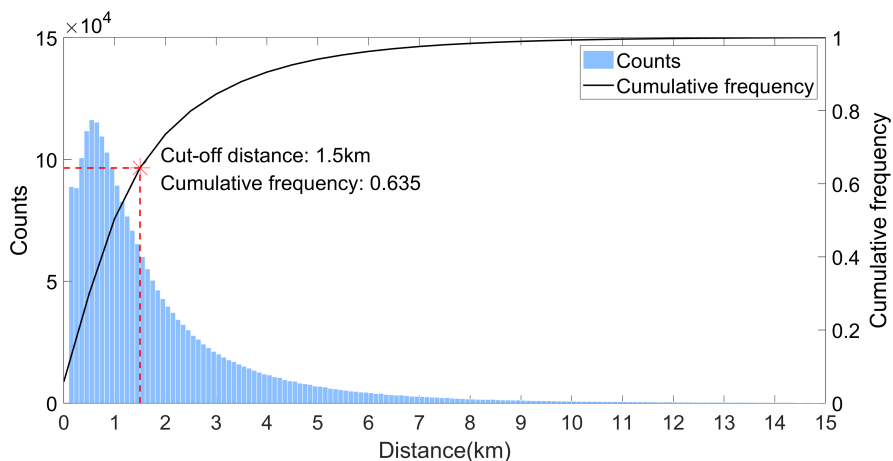

Figure 2 Distributions of bike-sharing trip lengths.

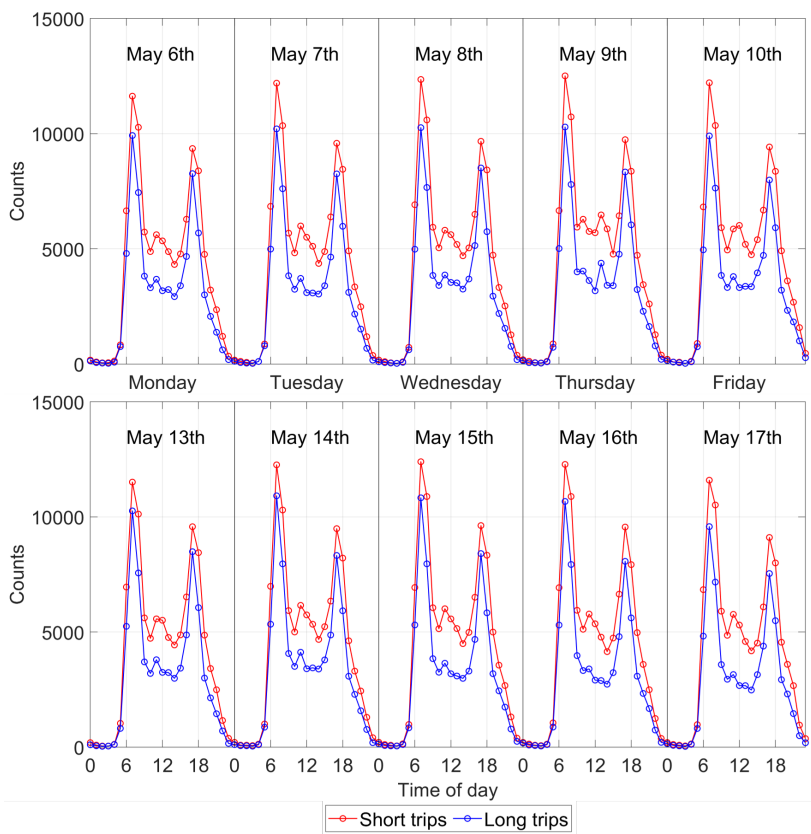

Figure 3 The hourly changes of bike-sharing demand.

Table 1 Statistical descriptions of bike-sharing demand (dependent variables).

| Variable | Average **B**ike-sharing **D**emand in all grids (per hour) | Max | Min | Mean | S.D. |
| --- | --- | --- | --- | --- | --- |
| BDSM | **S**hort-length trips at **M**orning peaks | 253 | 10 | 48.95 | 40.94 |
| BDLM | **L**ong-length trips at **M**orning peaks | 168 | 10 | 35.97 | 28.17 |
| BDSO | **S**hort-length trips at **O**ff-peaks | 150 | 10 | 33.02 | 25.17 |
| BDLO | **L**ong-length trips at **O**ff-peaks | 87 | 10 | 22.33 | 13.54 |
| BDSE | **S**hort-length at **E**vening peaks | 211 | 10 | 41.14 | 33.01 |
| BDLE | **L**ong lengths at **E**vening peaks | 163 | 10 | 33.76 | 24.91 |

Three categories of independent variables are considered in this study. Regarding travel-related variables, the number of metro stations and bus stops and the accessibility to two services are used to measure public transit services. The road length in grids is also considered. As for socio-economic characteristics, the number of households and housing prices are also treated as independent variables. The number of households can better represent the population size than the number of dwellings, which is commonly used in the literature. Housing prices can reflect the consuming capability to some extent. Built environment is measured using points of interest (POIs) owing to the unavailability of land use data. The housing prices and number of households are obtained from Lianjia (the largest real estate agency in China), and the POIs are obtained from AMAP.com (one of the largest location-based services providers in China). The independent variables are calculated in ArcGIS 10.6. The statistical descriptions of independent variables are shown in Table 2.

Table 2 Statistical descriptions of independent variables.

| Independent variable | Description | Max | Min | Mean | S.D. |
|---|---|---|---|---|---|
| Travel-related variables | | | | | |
| Number of metro stations | The number of metro stations in the grid | 1 | 0 | 0.09 | 0.29 |
| Metro accessibility | Distance from the grid center to the closest metro station (unit: m) | 4756.84 | 0.76 | 1533.12 | 961.05 |
| Number of bus stops | The number of bus stops in the grid | 11 | 0 | 1.23 | 1.31 |
| Bus accessibility | Distance from the grid center to the closest bus stop (unit: m) | 4016.55 | 0.78 | 458.02 | 451.02 |
| Road lengths | The road length in the grid (unit: m) | 5854.69 | 0 | 1080.26 | 827.09 |
| Built environment-related variables | | | | | |
| Restaurants | The number of restaurants in the grid | 153 | 0 | 9.4 | 13.36 |
| Firms | The number of firms in the grid | 495 | 0 | 14.77 | 26.01 |
| Education facilities | The number of education facilities in the grid | 87 | 0 | 4.94 | 7.76 |
| Banks | The number of banks in the grid | 42 | 0 | 3.46 | 5.7 |
| Government agencies | The number of government agencies in the grid | 73 | 0 | 6.88 | 8.44 |
| Hospitals | The number of hospitals in the grid | 39 | 0 | 4.86 | 5.91 |
| Hotels | The number of hotels in the grid | 115 | 0 | 3.63 | 10.42 |
| Gyms | The number of gyms in the grid | 59 | 0 | 5.33 | 6.66 |
| Service facilities | The number of service facilities in the grid | 113 | 0 | 9.66 | 11.66 |
| Scenic spots | The number of scenic spots in the grid | 41 | 0 | 0.77 | 2.63 |
| Shops | The number of shops in the grid | 142 | 0 | 7.26 | 10.78 |
| Socio-economic characteristics | | | | | |
| Number of households | The number of households in the grid | 15598 | 0 | 2214.87 | 2283.92 |
| Housing prices | Average housing prices in the grid (unit: yuan) | 78094.94 | 8101 | 26273.81 | 9340.05 |

# 4 Methodology

## 4.1 Tests of multicollinearity

The multicollinearity among independent variables would make the estimated results biased. Thus, the independent variables that are highly correlated with other variables should be removed. In this study, the inflation variance factor (VIF) scores are applied to measure the multicollinearity. The VIF scores higher than 10 are removed.

## 4.2 Spatial autocorrelation

Tests of spatial autocorrelation are conducted before using spatial regression models. The Moran's I test is a method that is commonly used for testing spatial autocorrelations. It is also used in this study and the Moran index can be obtained as follows.

$$I = \frac{N}{\sum_{i=1}^{N}\sum_{j=1}^{N}w_{ij,k}} \cdot \frac{\sum_{i=1}^{N}\sum_{j=1}^{N}w_{ij,k}(x_{i,k}-\bar{x}_k)(x_{j,k}-\bar{x}_k)}{\sum_{i=1}^{N}(x_{i,k}-\bar{x}_k)^2} (i \neq j) \tag{1}$$

where $N$ is the number of grids; $w_{ij}$ is the spatial weight between grid $i$ and grid $j$; $x_{i,k}$ and $x_{j,k}$ are the number of independent variable $k$ in grid $i$ and grid $j$, respectively; $\bar{x}_k$ is the average number of independent variable $k$.

The value of the Moran's I index is between -1 and 1. A value bigger than 0 indicates a positive spatial correlation and vice versa. Additionally, a higher absolute value means a more significant correlation. When the value is equal to 0, the variable is randomly distributed.

## 4.3 Spatial regression model

As shown in Figure 1, the bike-sharing demand is unevenly distributed and mainly concentrates around the urban central areas owing to the urban form and function. The bike-sharing demand is positively skewed to the right, which is contrary to the assumption of normal distribution in the ordinary least squares regression model. The log transformation is commonly used to address this issue in the literature (Qian and Ukkusuri, 2015). However, the data with log transformation cannot meet the assumption of normal distribution in this study, as shown in Figure 4. Thus, it is assumed that the bike-sharing demand follows the Poisson distribution (Chen et al., 2021). A GWPR model is applied in this study, which can accommodate the spatial autocorrelations. It should be noted that we remove the grids with average demand less than 1, because the demand cannot be a decimal.

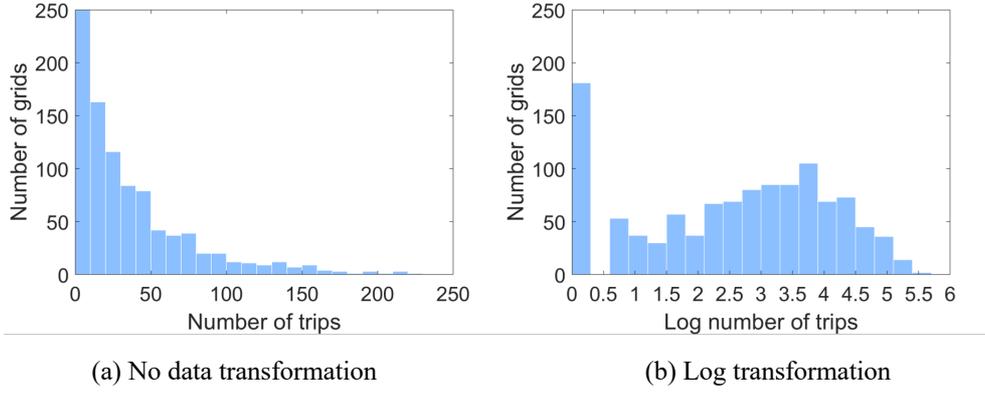

(a) No data transformation      (b) Log transformation

Figure 4 Bike-sharing demand with and without transformation.

The GWPR model is written as follows.

$$O_i^t \sim Poisson\left[E_i^t \exp\left(\sum_k \beta_k^t(u_i^t, v_i^t) x_{k,i}^t + \beta_0^t(u_i^t, v_i^t)\right)\right] \quad (2)$$

where $(u_i^t, v_i^t)$ is the center coordinate of grid $i$; $O_i^t$ is the bike-sharing demand in grid $i$; $x_{k,i}^t$ is the number of independent variable $k$ in grid $i$; $\beta_0^t(u_i^t, v_i^t)$ is the intercept term; $E_i^t$ is the offset variable. In this study, $E_i^t$ is equal to 1. $\beta_k^t(u_i^t, v_i^t)$ can be obtained by maximizing the log likelihood of the geographically weighted equation as follows (Nakaya et al., 2005).

$$\max L(u_i^t, v_i^t) = \sum_j^N \left(-\widehat{O_j^t}\left(\beta(u_i^t, v_i^t)\right) + O_j^t \log \widehat{O_j^t}\left(\beta(u_j^t, v_j^t)\right)\right) \cdot w_{ij}^t(\|(u_i^t, v_i^t) - (u_j^t, v_j^t)\|) \quad (3)$$

where $\widehat{O_j^t}(\beta(u_i^t, v_i^t)$ is the predicted number of bike-sharing demand in grid $j$ based on the demand in grid $i$, which can be obtained as follows.

$$\widehat{O_j^t}(\beta(u_i^t, v_i^t) = E_j^t \exp\left(\beta_0^t(u_i^t, v_i^t) + \sum_k \beta_k^t(u_i^t, v_i^t) x_{k,j}^t\right) \quad (4)$$

where $w_{ij}^t$ is the spatial weight between grid $i$ and grid $j$. The weight decreases as the distance between the two grids increases.

The function of adaptive bi-square kernel is used to define the spatial weight matrix as follows.

$$w_{ij}^t = \begin{cases} \left(1 - (\|(u_i^t, v_i^t) - (u_j^t, v_j^t)\|)^2\right)/(\theta_{i(n)}^t)^2 & \|(u_i^t, v_i^t) - (u_j^t, v_j^t)\| < \theta_{i(n)}^t \\ 0 & otherwise \end{cases} \quad (5)$$

where $\theta_{i(k)}^t$ is the nearest distance between grid $i$ and its $k$th neighborhood grid.

Considering that both global and local variables are included in the model, the GWPR model is further extended to an SGWPR model as follows.

$$O_i^t \sim Poisson\left[E_i^t \exp\left(\sum_k \beta_k^t(u_i^t, v_i^t) x_{k,i}^t + \sum_m \gamma_m^t z_{m,i}^t + \beta_0^t(u_i^t, v_i^t)\right)\right] \quad (6)$$

where $\gamma_m^t$ is the coefficient of the $m$th independent variable.

# 5 Results
## 5.1 Results of multicollinearity tests and spatial autocorrelations

The results of multicollinearity tests are shown in Table 3. All independent variables are kept as their VIF scores are smaller than 10.

Table 3 VIF scores of independent variables.

| Variables | VIF |
| --- | --- |
| Number of metro stations | 1.19 |
| Metro accessibility | 1.33 |
| Number of bus stops | 1.3 |
| Bus accessibility | 1.27 |
| Road lengths | 1.05 |
| Restaurants | 2.97 |
| Firms | 1.34 |
| Education facilities | 1.46 |
| Banks | 1.7 |
| Government agencies | 1.49 |
| Hospitals | 2.12 |
| Hotels | 1.31 |
| Gyms | 2.01 |
| Service facilities | 3.35 |
| Scenic spots | 1.26 |
| Shops | 1.64 |
| Number of households | 1.58 |
| Housing prices | 1.55 |

The Moran's I test result for variables are shown in Table 4. Except for the number of metro stations, the coefficients of most independent variables are significant at the 1% significance level. Additionally, the positive Z-scores indicate that the independent variables have clustering characteristics. Thus, it is necessary to consider the spatial heterogeneities based on a spatial regression model. As the Z-score of the number of metro stations is significantly smaller than 0, the number of metro stations is treated as a global variable and other variables are treated as local variables.

As shown in Figure 5, the local Moran's I index suggests that the bike-sharing demand shows significant spatial clustering characteristics. It is found that the bike-sharing demand mainly concentrates in District A, where land use is more mixed and cycling facilities are better deployed. In District B, some bike-sharing hotspots for short trips are identified, including the ZhongShanLu community and WangChuanChang community, which are known as residential areas. At off-peaks and evening peaks, the bike-sharing demand of long trips has a hotspot (i.e., XiHuDao community) in District D. These spatial clustering characteristics suggest that there exist spatiotemporal heterogeneities in the bike-sharing demand. In the following section, the spatiotemporal variations in the bike-sharing demand with different trip lengths will be further analyzed.

Table 4 Moran's I test results for variables.

| Variable | Moran's I | Z-score | P value | Expected Index |
|---|---|---|---|---|
| BDSM | 0.674 | 30.692 | 0 | -0.0009 |
| BDLM | 0.710 | 32.326 | 0 | -0.0009 |
| BDSO | 0.778 | 35.435 | 0 | -0.0009 |
| BDLO | 0.796 | 36.244 | 0 | -0.0009 |
| BDSE | 0.774 | 35.242 | 0 | -0.0009 |
| BDLE | 0.758 | 34.510 | 0 | -0.0009 |
| Number of metro stations | -0.029 | -1.268 | 0.205 | -0.0009 |
| Metro accessibility | 0.771 | 35.004 | 0 | -0.0009 |
| Number of bus stops | 0.269 | 12.287 | 0 | -0.0009 |
| Bus accessibility | 0.476 | 21.771 | 0 | -0.0009 |
| Road lengths | 0.668 | 30.390 | 0 | -0.0009 |
| Restaurants | 0.492 | 22.568 | 0 | -0.0009 |
| Firms | 0.409 | 19.693 | 0 | -0.0009 |
| Education facilities | 0.477 | 21.909 | 0 | -0.0009 |
| Banks | 0.445 | 20.308 | 0 | -0.0009 |
| Government agencies | 0.469 | 21.403 | 0 | -0.0009 |
| Hospitals | 0.484 | 22.028 | 0 | -0.0009 |
| Hotels | 0.404 | 18.764 | 0 | -0.0009 |
| Gyms | 0.440 | 20.061 | 0 | -0.0009 |
| Service facilities | 0.514 | 23.446 | 0 | -0.0009 |
| Scenic spots | 0.436 | 20.691 | 0 | -0.0009 |
| Shops | 0.292 | 13.532 | 0 | -0.0009 |
| Number of households | 0.312 | 14.187 | 0 | -0.0009 |
| Housing prices | 0.913 | 41.546 | 0 | -0.0009 |

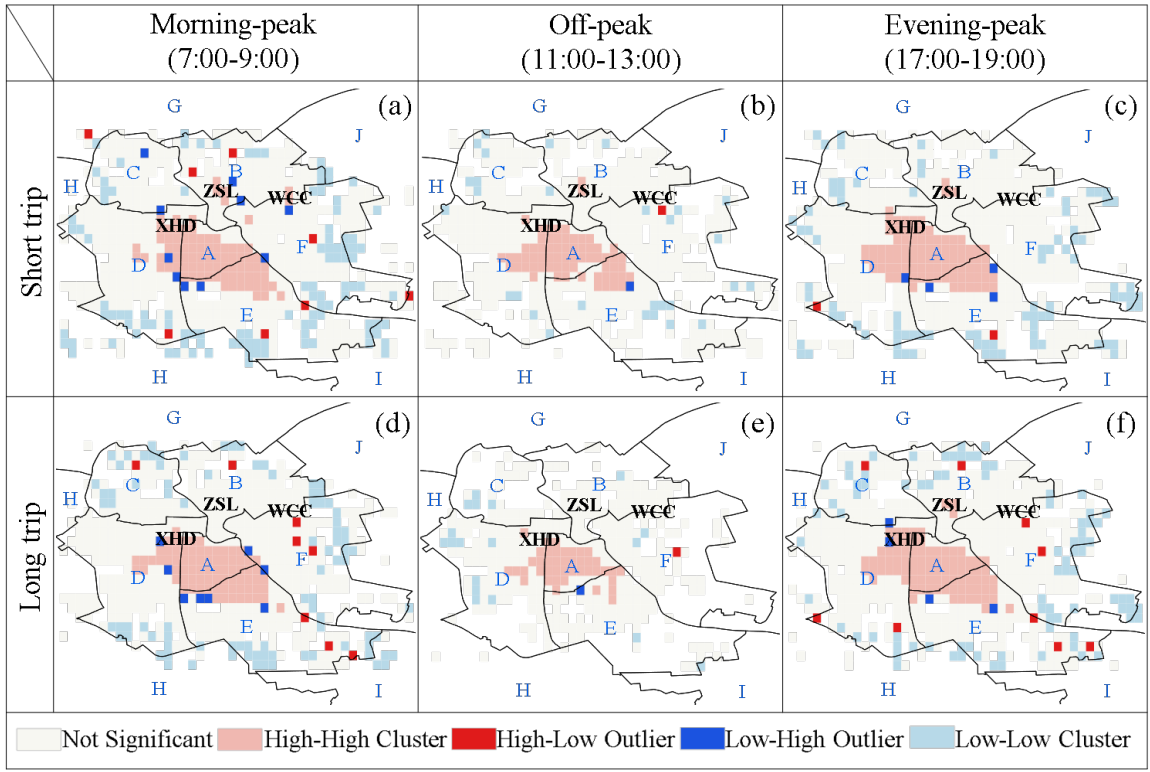

Figure 5 Spatial clustering characteristics of the bike-sharing demand (ZSL: ZhongShanLu; WCC: WangChuangChang; XHD: XiHuDao). *High-High Cluster* (*Low-Low Cluster*): the grid has high (low) bike-sharing demand and its neighborhoods also have high (low) bike-sharing demand; *High-Low outlier* (*Low-High outlier*): the grid has high (low) bike-sharing demand and its neighborhood grids have low (high) bike-sharing demand.

## 5.2 Estimated results of the SGWPR model

### 5.2.1 Model fitness and selection of trip length threshold

To confirm the effectiveness of the SGWPR model, a generalized linear (GL) model is used as a comparison. The Akaike information criterion (AIC), Bayesian Information Criterion (BIC), small sample bias corrected AIC (AICc) and percent Deviance (pDev) are used as indices to compare the model fitness of the GL model and the SGWPR model. As shown in Table 5, the values of AIC, BIC and AICc for the SGWPR model are smaller while the pDev value is bigger. These results suggest that the SGWPR model is more powerful in explaining the relationship between built environment and bike-sharing demand.

Table 5 Comparison between the GL model and SGWPR model.

| Index | Morning peak | | | | Off-peak | | | | Evening peak | | | |
| --- | --- | --- | --- | --- | --- | --- | --- | --- | --- | --- | --- | --- |
| | Short trips | | Long trips | | Short trips | | Long trips | | Short trips | | Long trips | |
| | GLM | SGWPR | GLM | SGWPR | GLM | SGWPR | GLM | SGWPR | GLM | SGWPR | GLM | SGWPR |
| AIC | 6508.08 | 2247.25 | 3573.39 | 1528.64 | 3132.92 | 1203.91 | 818.61 | 546.53 | 4350.73 | 1590.34 | 2832.90 | 1355.52 |
| AICc | 6509.20 | 2722.80 | 3574.66 | 1862.27 | 3134.35 | 1433.003 | 820.70 | 597.27 | 4351.96 | 1867.88 | 2834.16 | 1510.01 |
| BIC | 6594.63 | 3642.35 | 3657.43 | 2622.07 | 3214.81 | 2064.027 | 893.62 | 886.60 | 4435.37 | 2637.67 | 2917.06 | 2166.23 |
| pDev | 0.678 | 0.919 | 0.681 | 0.907 | 0.637 | 0.906 | 0.704 | 0.858 | 0.686 | 0.919 | 0.700 | 0.894 |

To examine the impacts of built environment on bike-sharing demand with trip lengths, a sensitivity analysis is conducted to identify an appropriate threshold to distinguish trip lengths The Ddev values of the SGWPR models with different thresholds are calculated. As shown in Figure 6, the average pDev value of long- and short-trip models is maximal when the threshold takes a value of 1.5 km. Several previous studies have found that most bike-sharing trips that are used as a feeder mode of metro when the cost is fewer than 10 minutes (Jiang et al., 2018; Lin et al., 2019). In our sample, the average speed of shared bikes is 9.2 km/h, and a traveler would ride for about 1.5 km in 10 minutes. Thus, most short-distance bike-sharing users may use shared bikes as a feeder mode. This inference may be partially confirmed by the results, in which the number of metro stations and metro accessibility show greater impacts on short trips than long trips.

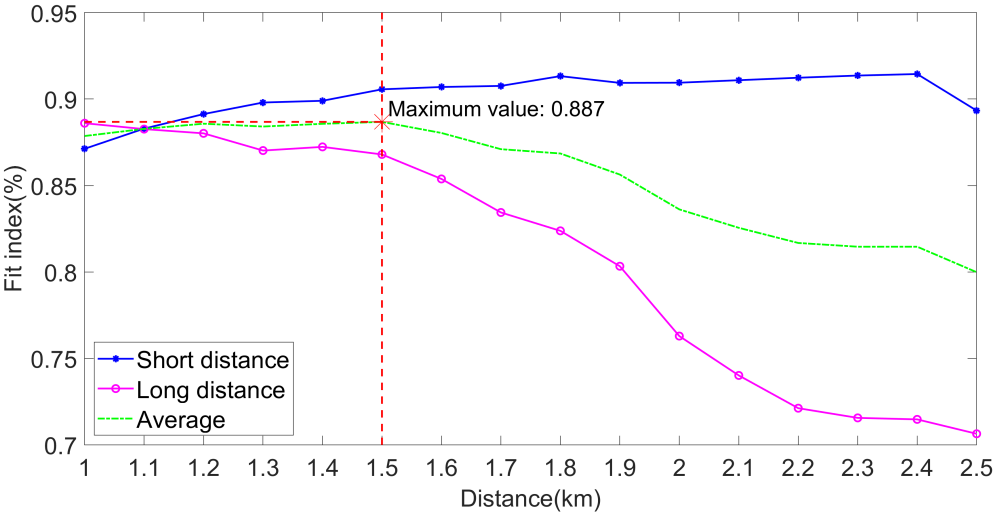

Figure 6 Sensitivity analysis of the thresholds of distinguishing trip lengths.

### 5.2.2 Impacts of built environment variables on bike-sharing demand

The estimated results of the SGWPR model are shown in Table 6. Owing to the spatial heterogeneities, the coefficients of independent variables vary across spatial areas. Considering the spatial non-stationarity, the min, lower quartile, mean, upper quartile and max values of the coefficients are used to measure the impacts of built environment on the bike-sharing demand. The results show that the number of bus stops and metro accessibility have positive and negative impacts on the short-distance bike-sharing demand at morning peaks, respectively. This result is reasonable because people usually choose to ride shared bikes for short distances from or to bus stops and metro stations. During off-peak hours, only restaurants have a positive impact on the short-distance bike-sharing demand, whereas the numbers of hotels, gyms, hospitals, households and firms show positive impacts on the long-distance bike-sharing demand. This may be explained by the reason that during spare time people are more likely to ride shared bikes for long distances than using motorized modes.

Table 6 Estimated results of the global variables in the SGWPR model.

| Variable | Morning-peak ||||||||||| Off-peak ||||||||||| Evening-peak |||||||||||
| | MIN || LQ || MEN || UQ || MAX || MIN || LQ || MEN || UQ || MAX || MIN || LQ || MEN || UQ || MAX ||
| | ST | LT | ST | LT | ST | LT | ST | LT | ST | LT | ST | LT | ST | LT | ST | LT | ST | LT | ST | LT | ST | LT | ST | LT | ST | LT | ST | LT | ST | LT |
|---|---|---|---|---|---|---|---|---|---|---|---|---|---|---|---|---|---|---|---|---|---|---|---|---|---|---|---|---|---|---|
| NM | 2.96 | 2.99 | 3.63 | 3.32 | 3.78 | 3.48 | 4.06 | 3.77 | 5.11 | 4.95 | 2.95 | 2.97 | 3.30 | 3.02 | 3.40 | 3.07 | 3.66 | 3.26 | 4.59 | 3.68 | 3.00 | 3.25 | 3.53 | 3.40 | 3.61 | 3.48 | 3.82 | 3.68 | 4.93 | 4.50 |
| MA | -0.53 | -0.35 | -0.20 | -0.10 | -0.09 | -0.01 | -0.02 | 0.08 | 0.27 | 0.27 | -0.27 | -0.11 | -0.12 | -0.02 | -0.01 | 0.03 | 0.07 | 0.07 | 0.60 | 0.18 | -0.37 | -0.27 | -0.16 | -0.06 | -0.06 | 0.04 | 0.01 | 0.09 | 0.42 | 0.29 |
| NBA | -0.11 | -0.13 | 0.01 | -0.02 | 0.06 | 0.02 | 0.10 | 0.06 | 0.32 | 0.26 | -0.26 | -0.05 | -0.01 | -0.02 | 0.03 | 0.00 | 0.07 | 0.03 | 0.17 | 0.06 | -0.13 | -0.08 | -0.01 | -0.02 | 0.03 | 0.01 | 0.06 | 0.03 | 0.15 | 0.15 |
| BA | -0.32 | -0.27 | -0.09 | -0.08 | -0.03 | -0.02 | 0.03 | 0.03 | 0.39 | 0.20 | -0.30 | -0.13 | -0.05 | -0.08 | 0.01 | -0.02 | 0.05 | 0.01 | 0.12 | 0.07 | -0.23 | -0.23 | -0.06 | -0.05 | 0.00 | -0.01 | 0.04 | 0.03 | 0.16 | 0.19 |
| RL | -0.65 | -0.43 | -0.01 | -0.01 | 0.11 | 0.07 | 0.23 | 0.19 | 1.34 | 1.37 | -0.38 | -0.08 | -0.04 | -0.03 | 0.03 | 0.00 | 0.15 | 0.12 | 1.14 | 0.75 | -0.41 | -0.25 | -0.05 | 0.01 | 0.02 | 0.03 | 0.13 | 0.12 | 1.26 | 0.97 |
| Res | -0.30 | -0.35 | -0.06 | -0.09 | 0.03 | 0.01 | 0.19 | 0.11 | 0.68 | 0.43 | -0.30 | -0.16 | 0.03 | 0.03 | 0.10 | 0.07 | 0.20 | 0.12 | 0.54 | 0.20 | -0.27 | -0.23 | 0.02 | -0.03 | 0.09 | 0.04 | 0.19 | 0.11 | 0.59 | 0.35 |
| Firms | -0.58 | -0.69 | -0.04 | 0.03 | 0.05 | 0.11 | 0.13 | 0.19 | 0.83 | 0.71 | -0.31 | -0.03 | -0.03 | 0.01 | 0.02 | 0.04 | 0.06 | 0.07 | 0.27 | 0.11 | -0.59 | -0.12 | -0.07 | 0.01 | -0.01 | 0.06 | 0.05 | 0.11 | 0.24 | 0.36 |
| EF | -0.15 | -0.25 | 0.05 | 0.01 | 0.15 | 0.11 | 0.25 | 0.20 | 0.79 | 0.62 | -0.16 | -0.01 | 0.03 | 0.05 | 0.11 | 0.08 | 0.18 | 0.13 | 0.49 | 0.19 | -0.19 | -0.14 | 0.06 | 0.03 | 0.12 | 0.07 | 0.21 | 0.13 | 0.47 | 0.24 |
| Banks | -0.28 | -0.28 | 0.04 | 0.06 | 0.08 | 0.10 | 0.15 | 0.16 | 0.39 | 0.58 | -0.04 | 0.03 | 0.06 | 0.05 | 0.09 | 0.07 | 0.12 | 0.10 | 0.30 | 0.15 | -0.09 | -0.09 | 0.04 | 0.03 | 0.08 | 0.06 | 0.12 | 0.10 | 0.26 | 0.23 |
| GA | -0.34 | -0.16 | 0.02 | 0.01 | 0.08 | 0.07 | 0.16 | 0.11 | 0.52 | 0.28 | -0.12 | -0.04 | 0.01 | 0.01 | 0.05 | 0.03 | 0.10 | 0.04 | 0.27 | 0.10 | -0.17 | -0.07 | 0.01 | 0.02 | 0.04 | 0.05 | 0.09 | 0.10 | 0.30 | 0.25 |
| Hospitals | -0.28 | -0.24 | -0.03 | 0.01 | 0.10 | 0.10 | 0.18 | 0.16 | 0.47 | 0.32 | -0.39 | -0.06 | -0.10 | 0.02 | 0.06 | 0.05 | 0.13 | 0.07 | 0.25 | 0.12 | -0.28 | -0.07 | -0.04 | 0.02 | 0.07 | 0.06 | 0.13 | 0.10 | 0.22 | 0.20 |
| Hotels | -1.80 | -0.86 | -0.02 | 0.01 | 0.05 | 0.06 | 0.18 | 0.20 | 1.00 | 0.93 | -0.41 | -0.11 | -0.02 | -0.03 | 0.02 | -0.01 | 0.06 | 0.02 | 0.47 | 0.09 | -0.94 | -0.21 | -0.02 | -0.03 | 0.02 | -0.01 | 0.07 | 0.04 | 0.59 | 0.18 |
| Gyms | -0.38 | -0.37 | -0.17 | -0.16 | -0.05 | -0.05 | 0.05 | 0.02 | 0.27 | 0.24 | -0.41 | -0.11 | -0.09 | 0.01 | 0.03 | 0.05 | 0.07 | 0.07 | 0.24 | 0.12 | -0.33 | -0.16 | -0.10 | -0.05 | 0.02 | 0.05 | 0.06 | 0.10 | 0.32 | 0.22 |
| SF | -0.37 | -0.37 | -0.06 | -0.07 | 0.03 | 0.00 | 0.10 | 0.05 | 0.45 | 0.28 | -0.25 | -0.13 | -0.02 | -0.08 | 0.04 | -0.02 | 0.11 | 0.04 | 0.40 | 0.14 | -0.44 | -0.14 | -0.04 | -0.05 | 0.04 | 0.01 | 0.11 | 0.06 | 0.28 | 0.13 |
| SS | -0.67 | -0.83 | -0.15 | -0.18 | -0.03 | -0.04 | 0.09 | 0.04 | 0.53 | 0.51 | -0.47 | -0.28 | -0.08 | -0.07 | -0.02 | -0.03 | 0.08 | 0.01 | 0.46 | 0.07 | -0.54 | -0.43 | -0.08 | -0.10 | 0.00 | -0.02 | 0.15 | 0.03 | 0.49 | 0.32 |
| Shops | -0.57 | -0.32 | -0.03 | -0.01 | 0.05 | 0.05 | 0.14 | 0.10 | 0.35 | 0.25 | -0.25 | -0.03 | 0.02 | 0.03 | 0.06 | 0.07 | 0.15 | 0.10 | 0.26 | 0.15 | -0.57 | -0.11 | 0.02 | 0.02 | 0.07 | 0.05 | 0.15 | 0.11 | 0.36 | 0.27 |
| NH | -0.23 | -0.21 | -0.01 | -0.02 | 0.06 | 0.03 | 0.13 | 0.09 | 0.32 | 0.22 | -0.08 | -0.03 | -0.01 | 0.02 | 0.05 | 0.05 | 0.12 | 0.07 | 0.26 | 0.10 | -0.02 | -0.08 | 0.05 | 0.06 | 0.09 | 0.10 | 0.15 | 0.15 | 0.28 | 0.24 |
| HP | -0.52 | -0.67 | 0.01 | 0.01 | 0.09 | 0.10 | 0.23 | 0.25 | 0.85 | 0.64 | -0.30 | 0.01 | 0.04 | 0.09 | 0.13 | 0.12 | 0.26 | 0.16 | 0.58 | 0.38 | -0.16 | -0.24 | 0.07 | 0.07 | 0.19 | 0.18 | 0.32 | 0.27 | 0.66 | 0.69 |

Note: ST: short trips; LT: long trips; NM: number of metro stations; MA: metro accessibility; NBS: Number of bus stops; BA Bus accessibility; RL: road lengths; EF: education facilities; GA: government agencies; SF: service facilities; SS: scenic spots; NH: number of households; HP: housing prices.

The impact of global variables on bike-sharing demand is shown in Table 7. The results suggest that the number of metro stations shows a positive impact on bike-sharing demand. Moreover, the number of metro stations shows greater impacts on short-distance bike-sharing demand than long-distance bike-sharing demand at morning peaks and off peaks. The closer relationships between metro and short-distance bike-sharing demand during these periods suggest that bike-sharing plays a more important role in complementing metro travels at morning peaks.

Table 7 Estimated results of the local variables in the SGWPR model.

| Variable | Morning peaks | | | | Off-peaks | | | | Evening peaks | | | |
|---|---|---|---|---|---|---|---|---|---|---|---|---|
| | Short trips | | Long trips | | Short trips | | Long trips | | Short trips | | Long trips | |
| | Coef. | Z value | Coef. | Z value | Coef. | Z value | Coef. | Z value | Coef. | Z value | Coef. | Z value |
| Number of metro stations | 0.212 | 31.851 | 0.185 | 22.212 | 0.111 | 11.848 | 0.078 | 5.963 | 0.131 | 17.325 | 0.137 | 16.923 |

### 5.2.3 Spatiotemporal heterogeneity analysis

To further analyze the spatiotemporal heterogeneity in bike-sharing demand, the impacts of key independent variables on bike-sharing demand are visually presented and explicitly analyzed as follows. Except for the impacts of the key variables on the bike-sharing demand with different lengths, we also calculate the coefficient differences between the results of the short-trip model and long-trip model at different periods.

The impacts of bus accessibility on the bike-sharing demand with different lengths are shown in Figures 7. Although the impacts of bus accessibility show similar tendencies overall, some interesting phenomena could also be observed.

- In Region 1, bus accessibility only shows a negative impact on the long-distance bike-sharing demand at off-peaks and evening peaks, whereas its impact on short-distance bike-sharing demand is always positive. This result may be explained by the reason that Region 1 contains the well-known tourism spot of Tianjin (i.e., WuDaDao cultural tourism area) with high public transportation accessibility, and thus people may ride for short distances for entertainment activities at off-peaks.

- In Region 2, bus accessibility shows a negative impact on long-distance bike-sharing demand. This result may be explained by the reason that poor bus accessibility in this area forces people to use long bike-sharing trips to replace transit trips in Region 2.

- In Region 3, bus accessibility always shows a positive impact on short-distance bike-sharing demand, whereas its impact on long-distance bike-sharing demand is only positive on evening peaks. This result may be explained by the reason that the intensive public transportation network makes the bus replace some feeder bike-sharing trips of metros.

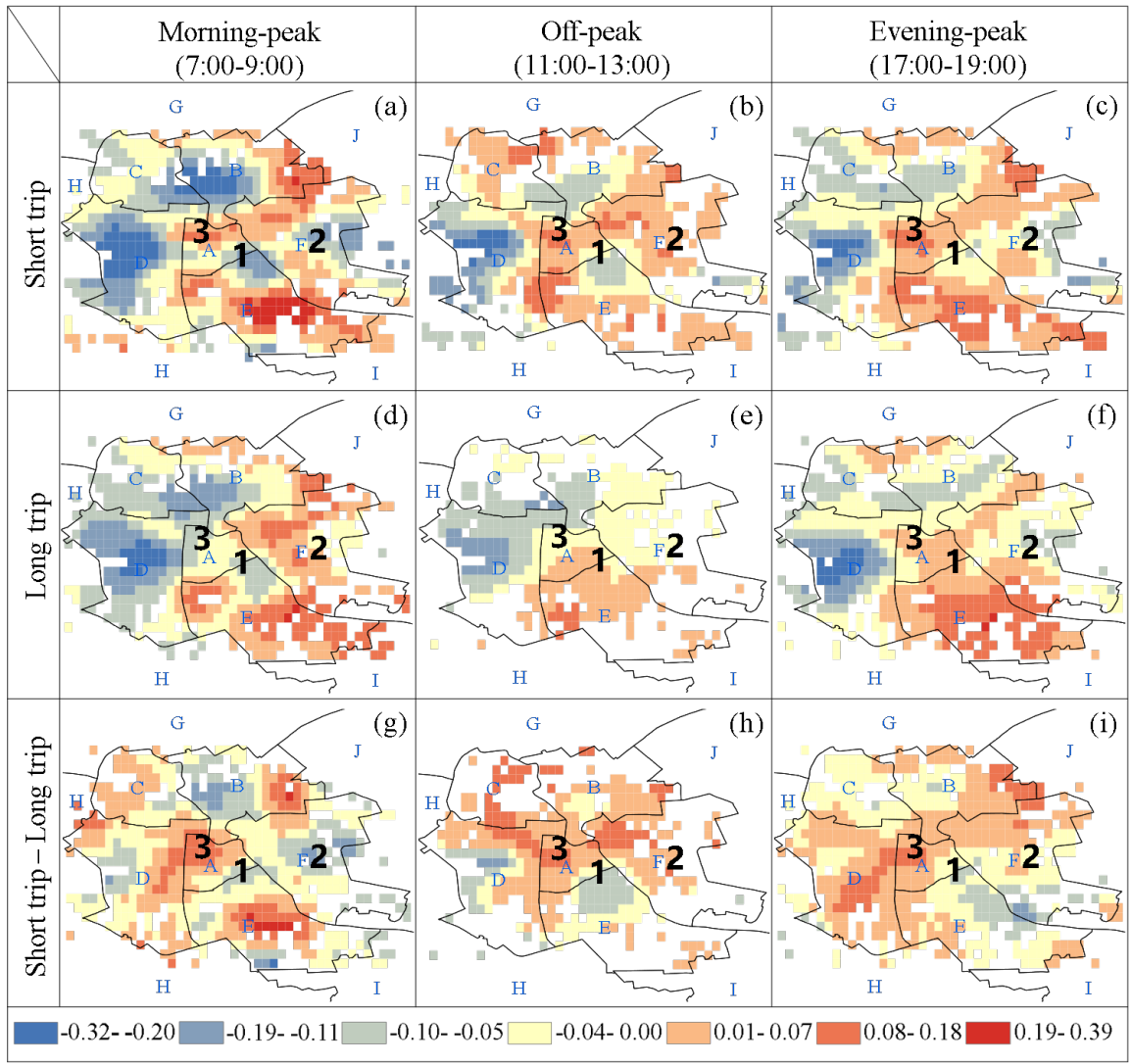

Figure 7 Spatiotemporal coefficients of bus accessibility, where "*Short trip - Long trip*" is the coefficient differences between the results of the short-trip model and long-trip model.

The impacts of the number of firms on the bike-sharing demand are shown in Figure 8. During peak hours, the number of firms shows positive impacts on bike-sharing demand. This relationship is rather significant in District A (i.e., *Heping* district), where is the working area of Tianjin and there are many firms and employment opportunities. The results suggest that the bike-sharing plays an important role in daily commuting trips; it is a new finding.

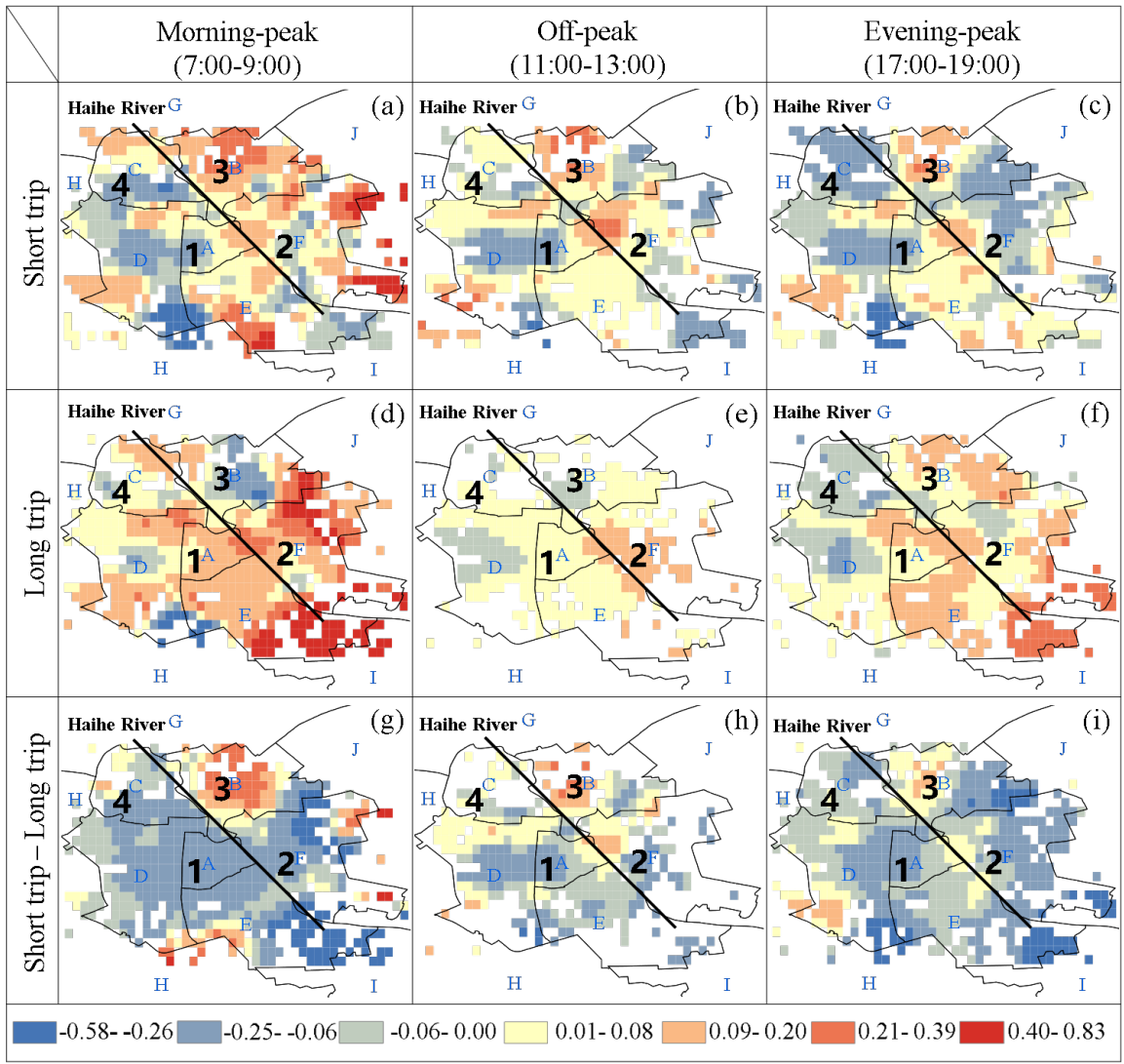

Figure 8 Spatiotemporal coefficients of the number of firms, where "*Short trip - Long trip*" is the coefficient differences between the results of the short-trip model and long-trip model.

- In Regions 2 and 4, the number of firms shows a negative impact on the short-distance bike-sharing demand during peak hours. This result may be explained by the reason that many industrial parks (with the features of large space but low working density in China) are in the areas with relatively poor public transit accessibility in the two regions. People who work in these industrial parks have to ride for long distances to reach their workplaces.

- In Region 1, the number of firms shows a negative impact on short-distance bike-sharing demand, whereas its impact on long-distance bike-sharing demand is positive. This result may be explained by the reason that the region is located near residential areas in urban core areas and there are only a few firms in the region. Better road connectivity encourages more bike-sharing trips with short lengths.

- In Region 3, the number of firms shows a negative impact on long-distance bike-sharing demand, although its land use is similar with Region 1. This result may be explained by the placement strategy of bike-sharing. Located in the north of *Haihe River*, Region 3 has lower economic development levels and fewer employment opportunities than Region 1. Thus, Region 1 is one of the most important placement areas. The results suggest that the placement strategy of bike-sharing has moderate effects on the relationship between built environment and bike-sharing demand and more bikes should be put into District B.

The impacts of housing prices on the bike-sharing demand are shown in Figure 9. In all models, the coefficients of housing prices show similar spatiotemporal patterns. Specifically, housing prices show mixed impacts on bike-sharing demand, and the impacts depend on comprehensive factors, including geographical locations and land use.

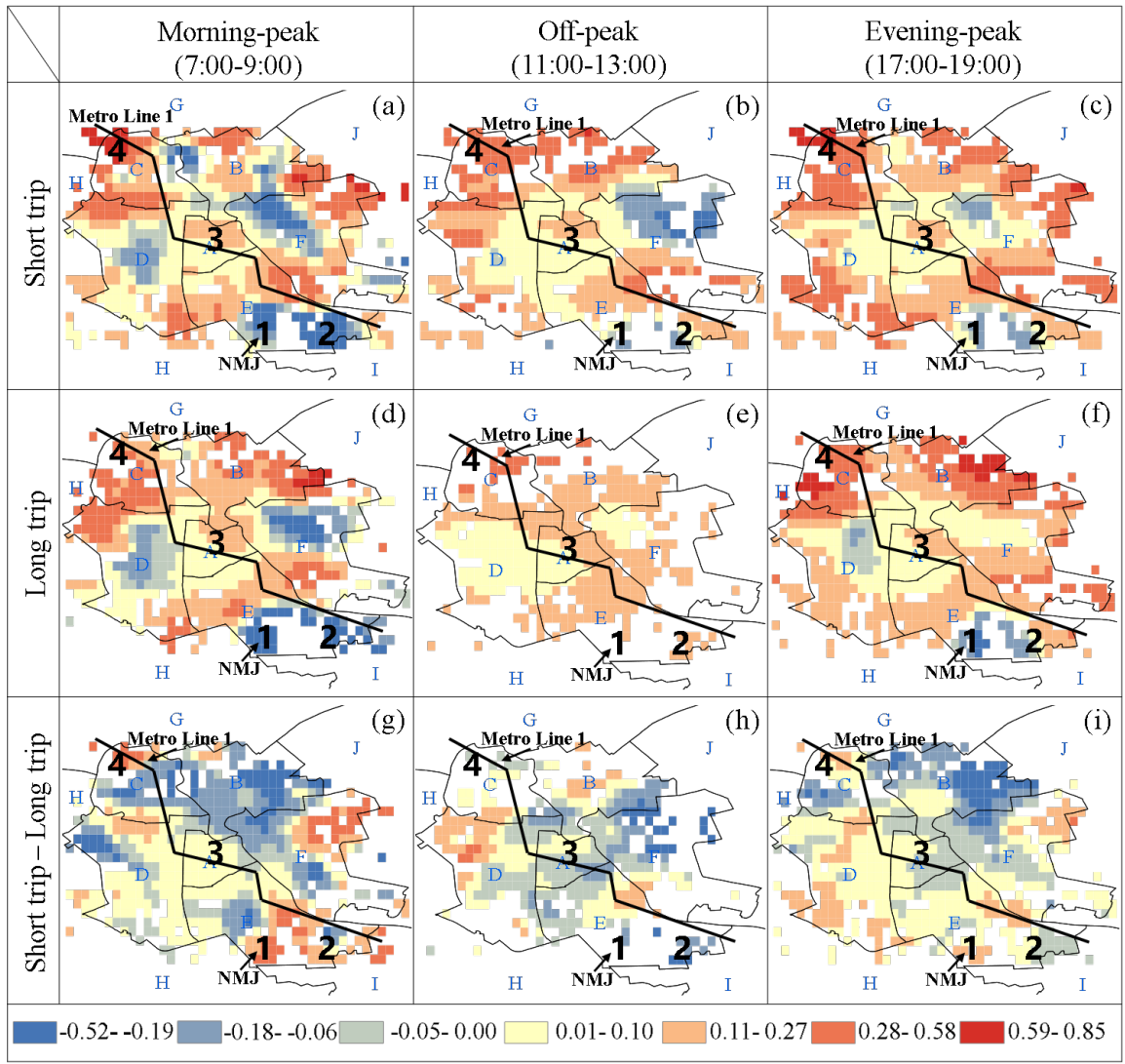

Figure 9 Spatiotemporal coefficients of housing prices (NMJ: New Meijiang), where "*Short trip - Long trip*" is the coefficient differences between the results of the short-trip model and long-trip model.

- In Region 1, many villas have been built in the "*New Meijiang*" area of Tianjin, where housing prices are relatively high. People living in these areas usually own private cars. The literature suggests that people would use a car once they have one (Wang et al., 2021), and thus people may generate fewer bike-sharing trips. Moreover, this area is vast and sparsely populated, which restricts the bike-sharing demand.
- In Region 3, the impact of housing prices shows a contrary sign with that in Region 1. This result may

be explained by the reason that Region 3 is in the core urban areas, which is the main placement area for bike-sharing. Additionally, the region is characterized by higher land use mix and better supporting facilities, which encourage bike-sharing trips.

- At morning peaks, housing prices show different impacts on the short- and long-distance bike-sharing demand in Region 2. This may be explained by the reason that multiple universities and metro stations in Region 2 cause high housing prices. At morning peaks, most students ride shared bikes for short distances to classrooms, and thus the results suggest that housing prices show a positive impact on short-distance bike-sharing demand.

- Region 4 is located around the end of *Metro Line 1* of Tianjin, which is characterized by low housing prices. Housing prices show positive impacts on the bike-sharing demand with different lengths. Moreover, the impact on short-distance bike-sharing demand is greater than long-distance bike-sharing demand, suggesting that the number of shared bikes is not enough in Region 4.

## 6 Discussion and conclusions

As a solution of first- and last-mile problems, bike-sharing has become the key segment to promote the green and low-carbon transformation of transportation. To better configure shared bikes, it is necessary to explore the impacts of built environment on the bike-sharing demand with different trip lengths and understand the spatiotemporal patterns of bike-sharing demand. Based on the bike-sharing GNSS trace data of Tianjin, China, this study employed an SGWPR model to explore the relationship between built environment and bike-sharing demand with different lengths. The spatiotemporal heterogeneity in the relationship is confirmed and several research and policy implications are provided.

The results of this study suggest that there are two peaks in the bike-sharing demand on weekdays. The long-distance bike-sharing demand concentrates in the core urban areas at peaks and off-peaks, whereas the short-distance bike-sharing demand is high in densely populated residential communities. Moreover, sensitivity analysis shows that 1.5 km is a reasonable threshold to divide the bike-sharing demand in terms of trip lengths. The results show that metro accessibility has a negative impact on the short-distance bike-sharing demand in most areas, indicating that most users treat bike-sharing as a feeder mode of metro (in Tianjin). The placement of shared bikes has moderating effects on the relationship between built environment and bike-sharing demand. Buses will attract some short-distance bike-sharing demand in the areas with better public transit accessibility, whereas people have to ride for long distances to replace motorized travels in the areas with poorer public transit accessibility. People tend to use bike-sharing for long trips in industrial areas whereas those living in core urban areas have more short-distance bike-sharing demand during peak hours. Although housing prices show similar impacts on bike-sharing demand, the impacts depend on local land use and geographical locations.

The findings provide policy implications for bike-sharing operators and urban planners. First, urban planners should provide more cyclist-friendly environments in residential areas around urban core areas, where the bike-sharing demand is high. Some policies that promote bike-sharing usage (e.g., managing illegal occupation of non-motorized lanes by cars and adding bicycle signal lights at large intersections) should be encouraged. Second, although there is potential bike-sharing demand in areas with concentrated industrial enterprises and areas along the metro lines, the number of bike-sharing places is few. Increasing the placements of bike-sharing in these areas may promote the proportion of green travels. Moreover, people tend to use cars for long trips in the areas with poor public transit areas. Increasing the construction of bike lanes and providing more timely bike-sharing placement services are important to encourage people to switch from cars to bikes. Finally, owing to the spatiotemporal heterogeneity in the relationship between built environment and bike-sharing demand, it is important to coordinate between land use and public transportation provision to promote the green travels of bike-sharing.

Several limitations of this study are also noticed. First, as individual characteristics (e.g., age and gender) are not contained in the dataset, the impacts of individual characteristics are not controlled in this study. Second, only bike-sharing trace data within two weeks in Tianjin are analyzed in this study. The spatial and temporal coverages of the data are expected to extend in future because the bike-sharing demand is influenced by local temperatures and topographies.

# Acknowledgement

This work was sponsored by the National Natural Science Foundation of China (71871010, 72204114).

# References

Barbour, N., Zhang, Y. and Mannering, F. A statistical analysis of bike sharing usage and its potential as an auto-trip substitute. Journal of Transport & Health 12, 253–262, 2019

Chen, C., Feng, T., Ding, C., Yu, B. and Yao, B. Examining the spatial-temporal relationship between urban built environment and taxi ridership: Results of a semi-parametric GWPR model. Journal of Transport Geography 96, 103172, 2021

Chang, X., Wu, J., He, Z., Li, D., Sun, H., & Wang, W. Understanding user's travel behavior and city region functions from station-free shared bike usage data. Transportation Research Part F: Traffic Psychology and Behaviour, 72, 81–95, 2020

El-Assi, W., Salah Mahmoud, M. and Nurul Habib, K. Effects of built environment and weather on bike sharing demand: a station level analysis of commercial bike sharing in Toronto. Transportation 44(3), 589–613, 2015

Eren, E. and Uz, V.E. A review on bike-sharing: The factors affecting bike-sharing demand. Sustainable Cities and Society 101882, 2019

Gao, K., Yang, Y., Li, A. and Qu, X. Spatial heterogeneity in distance decay of using bike sharing: An empirical large-scale analysis in Shanghai. Transportation Research Part D: Transport and Environment 94, 102814, 2021

Guo, Y. and He, S.Y. Built environment effects on the integration of dockless bike-sharing and the metro. Transportation Research Part D: Transport and Environment 83, 102335, 2020

Guo, Y., Zhou, J., Wu, Y. and Li, Z. Identifying the factors affecting bike-sharing usage and degree of satisfaction in Ningbo, China. PLOS ONE 12(9), 0185100, 2017

Ji, Y., Ma, X., He, M., Jin, Y. and Yuan, Y. Comparison of usage regularity and its determinants between docked and dockless bike-sharing systems: A case study in Nanjing, China. Journal of Cleaner Production 255, 120110, 2020

Ji, Y., Ma, X., Yang, M., Jin, Y. and Gao, L. Exploring Spatially Varying Influences on Metro-Bikeshare Transfer: A Geographically Weighted Poisson Regression Approach. Sustainability 10(5), 1526, 2018

Jiang, Y., Chen, X, Xu, X., Qiao, Y. Exploring the catchment area of public bike connecting to subway. Journal of Transportation Systems Engineering and Information Technology 18.S1:94-102, 2018

Levy, N., Golani, C. and Ben-Elia, E. An exploratory study of spatial patterns of cycling in Tel Aviv using passively generated bike-sharing data. Journal of Transport Geography 76, 325–334, 2019

Li, A., Zhao, P., Huang, Y., Gao, K. and Axhausen, K.W. An empirical analysis of dockless bike-sharing utilization and its explanatory factors: Case study from Shanghai, China. Journal of Transport Geography 88, 102828, 2020

Li, W., Chen, S., Dong, J. and Wu, J. Exploring the spatial variations of transfer distances between dockless bike-sharing systems and metros. Journal of Transport Geography 92, 103032, 2021

Lin, D., Zhang, Y., Zhu, R. and Meng, L. The analysis of catchment areas of metro stations using trajectory data generated by dockless shared bikes. Sustainable Cities and Society 49, 101598, 2019

Liu, H.-C. and Lin, J.-J. Associations of built environments with spatiotemporal patterns of public bicycle use. Journal of Transport Geography 74, 299–312, 2019

Lu, W., Scott, D.M. and Dalumpines, R. Understanding bike share cyclist route choice using GPS data: Comparing dominant routes and shortest paths. Journal of Transport Geography 71, 172–181, 2018

Ma, X., Ji, Y., Yuan, Y., Van Oort, N., Jin, Y. and Hoogendoorn, S. A comparison in travel patterns and determinants of user demand between docked and dockless bike-sharing systems using multi-sourced data. Transportation Research Part A 139, 148–173, 2020

Martin, E.W. and Shaheen, S.A. Evaluating public transit modal shift dynamics in response to bikesharing: a tale of two U.S. cities. Journal of Transport Geography 41, 315–324, 2014

Mohanty, S., Bansal, S. and Bairwa, K. Effect of integration of bicyclists and pedestrians with transit in New Delhi. Transport Policy 57, 31–40, 2017

Nakaya, T., Fotheringham, A.S., Brunsdon, C. and Charlton, M. Geographically weighted Poisson regression for disease association mapping. Statistics in Medicine 24(17), 2695–2717, 2005

Qian, X. and Ukkusuri, S.V. Spatial variation of the urban taxi ridership using GPS data. Applied Geography 59, 31–42, 2015

Radzimski, A. and Dzięcielski, M. Exploring the relationship between bike-sharing and public transport in Poznań, Poland. Transportation Research Part A: Policy and Practice 145, 189–202, 2021

Shen, Y., Zhang, X. and Zhao, J. Understanding the usage of dockless bike sharing in Singapore. International Journal of Sustainable Transportation 12(9), 686–700, 2018

Wang, X., Shao, C., Yin, C. and Dong, C. Exploring the effects of the built environment on commuting mode choice in neighborhoods near public transit stations: evidence from China. Transportation Planning and

Technology 44(1), 111–127, 2020a

Wang, Y., Zhan, Z., Mi, Y., Sobhani, A. and Zhou, H. Nonlinear effects of factors on dockless bike-sharing usage considering grid-based spatiotemporal heterogeneity. Transportation Research Part D: Transport and Environment 104, 103194, 2022

Wang, Z., Cheng, L., Li, Y. and Li, Z. Spatiotemporal Characteristics of Bike-Sharing Usage around Rail Transit Stations: Evidence from Beijing, China. Sustainability 12(4), 1299, 2020b

Wu, C., Kim, I. and Chung, H. The effects of built environment spatial variation on bike-sharing usage: A case study of Suzhou, China. Cities 110, 103063, 2021

Xin, F., Chen, Y., Wang, X. and Chen, X. Cyclist Satisfaction Evaluation Model for Free-Floating Bike-Sharing System: A Case Study of Shanghai. Transportation Research Record: Journal of the Transportation Research Board 2672(31), 21–32, 2018

Yang, Y., Heppenstall, A., Turner, A. and Comber, A. A spatiotemporal and graph-based analysis of dockless bike sharing patterns to understand urban flows over the last mile. Computers, Environment and Urban Systems 77, 101361, 2019

Yu, S., Liu, G. and Yin, C. Understanding spatial-temporal travel demand of free-floating bike sharing connecting with metro stations. Sustainable Cities and Society 74, 103162, 2021